*((Catch Phrase))*

DOI:

## Supramolecular Stacking of Doxorubicin on Carbon Nanotubes for *in vivo* Cancer Therapy**

*Zhuang Liu[1, 4]†, Alice C. Fan[2]†, Kavya Rakhra[2], Sarah Sherlock[1], Andrew Goodwin[1], Xiaoyuan Chen[3], Qiwei Yang[2], Dean W. Felsher[2]\*and Hongjie Dai[1]\**

Doxorubicin (DOX) is a member of the anthracycline class of chemotherapeutic agents used for the treatment of many common human cancers including aggressive non-Hodgkin's lymphoma [1, 2]. However, DOX is highly toxic in humans resulting in severe suppression of hematopoiesis, gastrointestinal toxicity [3], and cardiac toxicity [4]. To date, several approaches including its delivery using liposomes (DOXIL) [5] have been developed to reduce the toxicity and enhance the clinical utility of this highly active antineoplastic agent.

Carbon nanotubes have unique mechanical, optical and chemical properties with broad potential biomedical applications including imaging and cancer therapeutics [6-13]. Carbon nanotubes have been used as novel delivery vehicles *in vitro* to effectively shuttle various bio-molecules including drugs [13, 14], plasmid DNA [15] and small interfering RNA (siRNA) [16, 17] into cells *via* endocytosis [18]. On the other hand, the unique optical properties of single walled carbon nanotubes (SWNTs) have been utilized for biological imaging [7-11] as well as cancer cell destruction [19]. Various groups have studied the behaviors of this material in animal models [20-23]. Well functionalized SWNTs have been found to be non-toxic in mice over several months and can be gradually excreted by the biliary pathway from reticuloendothelial systems (RES) into feces, with the majority of SWNTs cleared within two months [20, 21]. Here, we explored the possibility of using supramolecular π-π stacking to load a cancer chemotherapy agent, doxorubicin (DOX), onto branched polyethylene glycol (PEG) functionalized SWNTs for *in vivo* drug delivery applications. We found that our new drug formulation, SWNT-DOX, afforded significantly enhanced therapeutic efficacy and a marked reduction in toxicity compared to free DOX and DOXIL.

It has been found that the surface of PEGylated SWNTs could be efficiently loaded with DOX via supramolecular π-π stacking (Fig. 1a) [13]. For Hipco SWNTs used in this work, the binding energy of DOX on nanotubes is estimated to be ~48 kJ/mol in water [13]. Branched PEG was used here to achieve prolonged blood circulation time [21]. The SWNT-DOX complex had an average length of 100 nm and diameter of 2~3 nm as determined using an atomic force microscope (AFM) (Fig. 1b). The loading ratio of DOX was determined by UV-VIS-NIR absorption spectra (Fig. 1c). Depending on solution pH and DOX concentration, DOX loading on nanotubes ranged from 1 to 4 grams DOX per gram of SWNTs [13]. We chose a loading of ~2.5 gram DOX per gram of SWNTs for the experiments in this work. We found that SWNT-DOX remained cytotoxic to cancer cells *in vitro* (Supplementary Fig. S1), likely due to the release of DOX from nanotubes inside cell endosomes and lysosomes with acidic pH [13].

In order to investigate the *in vivo* pharmacokinetics and biodistribution, SWNT-DOX was injected *via* tail vein into SCID mice bearing Raji lymphoma xenografts. Blood was drawn at different time points post injection with DOX concentrations measured by fluorescence spectra following published protocols [24]. After being loaded onto nanotubes, the DOX circulation half-life increased from 0.21 hours for free DOX to 2.22 hours for the SWNT-DOX formulation while the total area under curve ($AUC_{0-\infty}$) also increased from 5.3 mg·h/L to 78.8 mg·h/L (Fig. 2a). Mice were sacrificed at 6 hours post injection with major organs harvested and homogenized for DOX extraction [24]. The concentration of DOX in each organ was measured by fluorescence intensity. Free DOX tumor uptake was 0.68 percent of injected dose per gram tissue (%ID/g). Tumor uptake of DOX doubled to 1.51 %ID/g for SWNT-DOX, which was likely due to the prolonged circulation of SWNT-DOX that facilitated the enhanced permeability and retention (EPR) effect observed with various nanomaterials (Fig. 2b) [25]. As expected, in the SWNT-DOX formulation, DOX accumulated in the RES including liver and spleen. Despite the high RES uptake, SWNT-DOX did not show obvious hepatic toxicity from histology studies (Supplementary Fig. S4). The combined biodistribution information of SWNTs determined by Raman scattering and DOX detected by fluorescence (Supplementary Fig. S2) indicates that although most SWNT-DOX is still in the associated form in the first few hours after administration, the loaded DOX slowly dissociates from nanotubes and is excreted via kidneys. Most nanotubes are too large to be excreted by the kidneys and instead are slowly excreted via the biliary system into the feces [21].

The *in vivo* therapeutic efficacy of the SWNT-DOX complex was also investigated. SCID mice bearing Raji lymphoma xenograft tumors were treated weekly with SWNT-DOX, free DOX, or DOXIL. Sizes of subcutaneous tumors were measured over two weeks. Tumors in untreated controls rapidly increased by 7.53 ± 0.99 fold. The SWNT-DOX (5 mg/kg) treated group showed a greater inhibition of tumor growth than free DOX at the equivalent

[*]  1) Zhuang Liu, Sarah Sherlock, Andrew Gooding, Hongjie Dai∗
Department of Chemistry
Stanford University
Stanford, CA, 94305, USA
E-mail: Hongjie Dai (hdai@stanford.edu)

2) Alice Fan, Kavya Rakhra, Dean Felsher∗
Division of Oncology
Stanford University School of Medicine
Stanford, CA, 94305, USA
E-mail: Dean Felsher (dfelsher@stanford.edu)

3) Xiaoyuan Chen
Department of Radiology
Stanford University School of Medicine
Stanford, CA, 94305, USA

4) Zhuang Liu
Functional Nano & Soft Materials Laboratory (FUNSOM)
Soochow University
Suzhou, Jiangsu, 215123, China

[**] This work was supported, in part, by the National Institutes of Health (NIH) - National Cancer Institution (NCI) grants R01 CA135109-01 (HD), 1R01 CA89305-01A1 (DWF), 3R01 CA89305-0351 (DWF), 1R01 CA105102 Lymphoma Program Project (DWF), NIH-NCI In Vivo Cellular and Molecular Imaging Center Grant P50 (DWF), NIH-NCI Center for Cancer Nanotechnology Excellence Focused on Therapeutic Response at Stanford (HD), Burroughs Welcome Fund (DMF), the Damon Runyon Foundation (DWF), the Leukemia and Lymphoma Society (DWF, ACF), a Stanford Bio-X Initiative Grant (HD, DWF), an Enscyse grant (HD) and a Stanford Graduate Fellowship (ZL).

[†] Drs. Z Liu and A Fan equally contributed to this work.

dose (2.15 ± 0.16 fold tumor growth *versus* 2.90 ± 0.19, respectively, *p = 0.016*) (Fig. 4a, 5a). Tumors in mice treated with plain SWNT increased by 6.44 ± 0.42 fold, (*versus* untreated, *p = 0.34*), indicating that SWNTs alone do not have therapeutic efficacy (Supplementary Fig. S3). Mice treated with DOXIL (5 mg/kg) developed severe treatment toxicity and mortality, as discussed below, but the DOXIL-treated mice that did remain alive at 14 days exhibited partial tumor regression to 0.88 ± 0.19 of original size. Thus, loading of DOX onto SWNT increased the therapeutic efficacy over free DOX, but not DOXIL.

To evaluate toxicity of SWNT-DOX formulation, we measured body weight of mice in each cohort. Mice treated with free DOX (5 mg/kg) and DOXIL (5 mg/kg) each exhibited a 19% and 17% decrease in weight within 2 weeks, respectively (Fig. 3b), and appeared to be both thinner and weaker after treatment (Fig. 4a, Supplementary Fig. S5). Treatment with DOX and DOXIL resulted in 20% and 40% mortality, respectively. In marked contrast, mice treated with SWNT-DOX had stable weight and no mortalities. In another control experiment, sequential administration of SWNT followed by DOX showed similar efficacy and toxicity as free DOX (Supplementary Fig. S3), suggesting that loading of DOX onto SWNT is essential for the treatment outcomes. We note that SCID mice used in our study appeared to exhibit more DOX toxicity than as described in literature for several other mouse strains including nude mice [5, 26]. Thus in our xenograft model of aggressive B cell lymphoma, loading of DOX onto SWNT significantly attenuated toxicity associated with free DOX and DOXIL.

We considered that the reduced toxicity of SWNT-DOX might allow us to administer higher doses of DOX. Indeed, SWNT-DOX administered at a 10 mg/kg dose improved treatment efficacy even further, with tumor size at 2 weeks only 1.64 ± 0.11 fold increased (*p = 0.018 versus* 5 mg/kg SWNT-DOX, p < 0.001 *versus* 5 mg/kg free DOX) (Fig. 3a). Furthermore, SWNT-DOX (10 mg/kg) treatment resulted in neither a significant decrease in body weight nor an increase in mortality. In marked contrast, 10 mg/kg free DOX was a uniformly lethal dose within 2 weeks of treatment. Mouse tissues from SWNT-DOX (10 mg/kg) did not exhibit gross microscopic gastrointestinal, cardiac, hepatic or renal toxicity (Fig. 4b, S4). In marked contrast, mice treated with DOX alone (5 mg/kg) demonstrated disruption of the intestinal lining consistent with gastrointestinal mucositis, exhibiting complete loss of columnar epithelial cells at the tips of villi, thus exposing the underlying connective tissue core of lymphatic capillaries, blood capillaries and smooth muscle (Fig. 4b). Therefore, by decreasing gastrointestinal toxicity, SWNT-DOX also improved the dose intensity that could be achieved in mice, further enhancing the treatment efficacy.

Finally, to compare overall clinical efficacy as defined by morbidity-free survival, Kaplan-Maier analysis was performed. In this analysis, mice in each cohort were scored either when tumor volume increased by 2-fold, when body weight decreased by 10%, or when they died during treatment. In this model, we found that SWNT-DOX formulation was superior to DOX alone or DOXIL (DOX *versus* SWNT-DOX 5mg/kg or 10mg/kg, *p< 0.001*; DOXIL *versus* SWNT-DOX 5mg/kg, *p=0.013*; DOXIL *versus* SWNT-DOX 10mg/kg, *p< 0.001*) (Fig. 4c). We conclude that supramolecular π-π stacking of a therapeutic agent onto SWNTs provides a novel method to efficiently deliver high dose chemotherapy with an increase in clinical efficacy.

Our strategy has critical advantages over a recently described approach in paclitaxel (PTX) delivery with SWNTs[12]. First, DOX is directly loaded on SWNT surface by simple non-covalent π-π stacking, achieving much higher drug loading capacity on nanotubes (up to 4g DOX / 1g SWNT), due to the ultra-high surface area of SWNTs and the nature of non-covalent loading that is not limited by the number of available functional groups. It has been previously estimated that ~70-80% of nanotube surface can be occupied by stacked DOX [13]. Our strategy can be easily extended to a wide range of lipophilic aromatic drugs including daunarubicin, gefitinib and camptothecin analogs [27]. Second, unlike SWNT-PTX conjugate, in which the hydrophobic PTX is exposed to outside environment and reduces the circulation half-life of the conjugate, DOX in the SWNT-DOX complex is stacked on nanotube sidewall and thus protected by long branched PEG coating, allowing more stable drug loading and significantly prolonged blood circulation.

There are several possible reasons why SWNT-DOX exhibits greater therapeutic efficacy and less toxicity than equimolar amounts of free DOX to treated mice. SWNT-DOX has a larger size that hampers its filtration through glomerulus, unlike free DOX, which is rapidly cleared out from blood circulation by renal excretion. The branched PEG coating attenuates the clearance of SWNT-DOX by macrophages in RES, causing SWNT-DOX to have a prolonged blood circulation half life [21], which allows repeated passing of drug conjugates through tumor vessels and increased tumor uptake by the EPR effect.

Our previous studies suggest that DOX loaded on SWNTs is stable at neutral pH but released in acidic environments [13]. The tumor microenvironment is slightly acidic [28], facilitating the dissociation of DOX from its SWNT carrier [13]. Normal organs and tissues, however, have a neutral pH, at which SWNT-DOX remains stable without releasing free DOX. SWNT-DOX that was detected in the intestinal tissues likely remained conjugated without releasing free DOX because epithelial cells that line the gastrointestinal tract maintain a neutral to alkaline environment [29]. Unlike DOX molecules which can freely diffuse out from blood vessels and reach intestinal tissues including epithelial intestinal lining, SWNT-DOX has reduced ability to undergo such diffusion due to its larger size that may limit access to intestinal tissues that are far from blood vessels in the intestine. This decrease in gastrointestinal epithelial toxicity is likely to contribute to the reduced toxicity of SWNT-DOX (Fig. 4b). However, the mechanism of our observed toxicity difference between SWNT-DOX and DOXIL remains to be explored (Fig. S5). Further studies are required to compare the clinical efficacy and toxic effects of SWNT-DOX and DOXIL at various doses.

In contrast to many other inorganic nanoparticles such as quantum dots that contain toxic heavy metals, carbon nanotubes are composed purely of carbon atoms that are relatively non-toxic. Compared with other traditional drug carriers such as polymers and liposomes, SWNTs have valuable features that allow both the ability to image cancer as well as to deliver a therapy. Near-infrared (NIR) fluorescence, Raman scattering or photo-acoustic contrast properties of SWNTs can be utilized for both *in vitro* and *in vivo* imaging [7-11]. The strong NIR optical absorption ability of SWNTs can be used for photothermal therapy [19], that could be combined with chemotherapy delivered by nanotubes for enhanced treatment efficacy. In addition to passive tumor targeting relying on the EPR effect, active *in vivo* tumor targeting of SWNTs has been achieved by conjugation of targeting peptides [23] or antibodies [22] to nanotubes. Thus carbon nanotubes have the potential to be a effective mechanism of drug delivery that could improve therapeutic efficacy and reduce drug related toxicities while simultaneously serving as an imaging modality for cancer.

## Experimental Section

**DOX loading on functionalized SWNTs.** Phospholipid-branched PEG (7 kDa) was synthesized as described earlier [21]. Raw Hipco SWNTs (0.2 mg/mL) were sonicated in a 0.2 mM solution phospholipid–branched PEG for 30 min with a cup-horn sonicator followed by centrifugation at 24,000 g for 6 h, yielding a suspension of SWNTs with non-covalent phospholipid–branched PEG coating in the supernatant [13, 16, 21]. Excess surfactant was removed by repeated filtration through a 100 kDa MWCO filter (Millipore) and extensive washing with water.

DOX loading onto PEGylated SWNTs was done by mixing 0.5 mM of DOX with the PEGylated SWNTs at a nanotube concentration of ~0.05 mg/ml (~300 nM) at pH 8 for overnight. Unbound excess DOX was removed by filtration through a 100 kDa filter and washed thoroughly with water until the filtrate became free of reddish color (corresponding to free DOX). The formed SWNT-DOX complex was characterized by UV-Vis-NIR absorbance spectra with a Cary-6000i spectrophotometer as described previously and stored at 4°C.

**In vivo circulation and biodistribution studies.** Blood circulation was measured by drawing ~15 μl blood from the tail vein of Raji tumor bearing SCID mice post injection of free DOX or SWNT-DOX. The blood samples were dissolved in a lysis buffer 1 (1% SDS, 1% Triton X-100, 40 mM Tris Acetate, 10 mM EDTA, 10 mM DTT) with brief sonication. The concentration of SWNTs in the blood was measured by a Raman method [21]. DOX measurement was carried out following the protocol previously reported with minor modification [24]. Briefly, DOX was extracted by incubating blood samples in 1 ml of 0.75 M HCl in isopropanol (IPA) at -20°C overnight. After centrifugation at 24,000 g for 15 minutes, fluorescence of the supernatant was measured using a fluorolog-3 fluorometer. Note that DOX loaded on SWNTs can be completely pulled off from nanotubes by the extraction solution with ~100% recovery of fluorescence (DOX fluorescence is quenched once loaded on nanotubes).

To study biodistribution, mice were sacrificed at 6 h post injection of free DOX or SWNT-DOX. The organs/tissues (0.1 – 0.2 g of each) were wet-weighed and homogenized in 0.5 ml of lysis buffer 2 (0.25M sucrose, 40 mM Tris Acetate, 10 mM EDTA) with a PowerGen homogenizer (Fisher Scientific). For DOX measurement, 200 μl of tissue lysate was mixed with 100 μl of 10% Titron X-100. After strong vortex, 1 ml of extraction solution (0.75 M HCl in IPA) was added and the samples were incubated at -20°C overnight. After centrifugation at 24,000 g for 15 min, fluorescence of the supernatant was measured.

**Treatment of in vivo lymphoma xenograft model.** SCID mice were subcutaneously injected with 10 million Raji cells. Treatment was initiated when the tumors reached a size of ~400 mm$^3$ (2-3 weeks after tumor inoculation). Tumor bearing mice were intravenously injected with different formulations of DOX including free DOX, SWNT, SWNT-DOX and DOXIL at 5 mg/kg of normalized DOX dose (or 10 mg/kg for SWNT-DOX) as well as related controls weekly. The tumor sizes were measured by calipers three times a week and volume was calculated according to the formula (tumor length) X (tumor width)$^2$/2. Relative tumor volumes were calculated as $V/V_0$ ($V_0$ was the tumor volume when the treatment was initiated). Mice were weighed with the relative body weights normalized to their initial weights. Tumors and organs were collected at the end of treatment, fixed in formalin, embedded in paraffin and sectioned using a microtome. Standard hematoxylin and eosin (H&E) staining was carried for histological examinations.

**Statistical analysis.** Quantitative data were expressed as mean ± standard errors of the mean (SEM). Means were compared using student's t-test. P values < 0.05 were considered statistically significant.

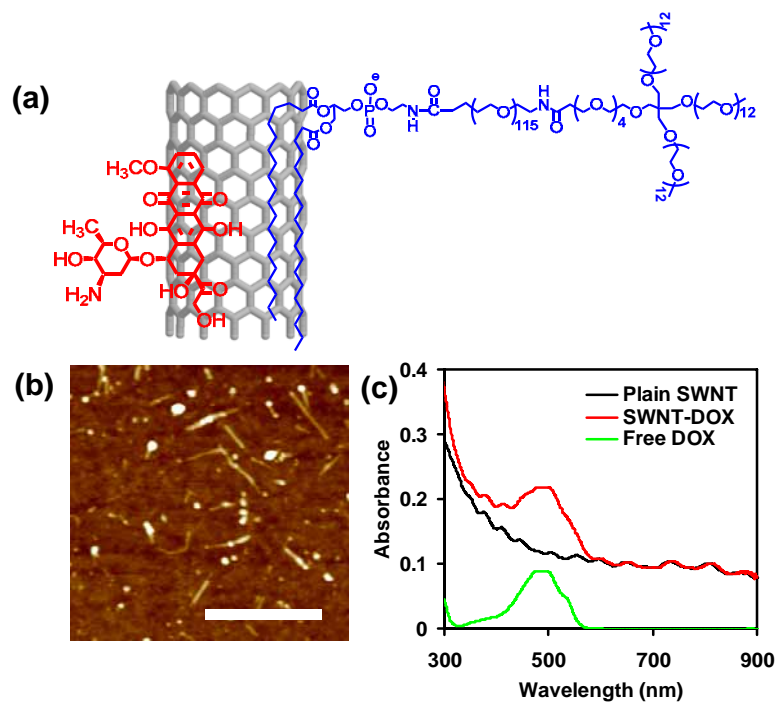

**Figure 1.** Doxorubicin was efficiently loaded on PEGylated SWNTs by π-π stacking. **(a)** A schematic drawing of SWNT-DOX complex. SWNTs were coated with biocompatible branched PEG with hydrophobic phospholipid anchored on the nanotube surface. DOX was densely packed on the aromatic nanotube surface via supramolecular π-π stacking. **(b)** An atomic force microscope (AFM) image of SWNT-DOX complexes. SWNT-DOX have average length of ~100 nm and diameter of 2~3 nm. Scale bar: 250 nm. **(c)** UV-VIS-NIR spectra of plain SWNT, SWNT-DOX and free DOX. The DOX loading was determined by the absorption peak at 490 nm.

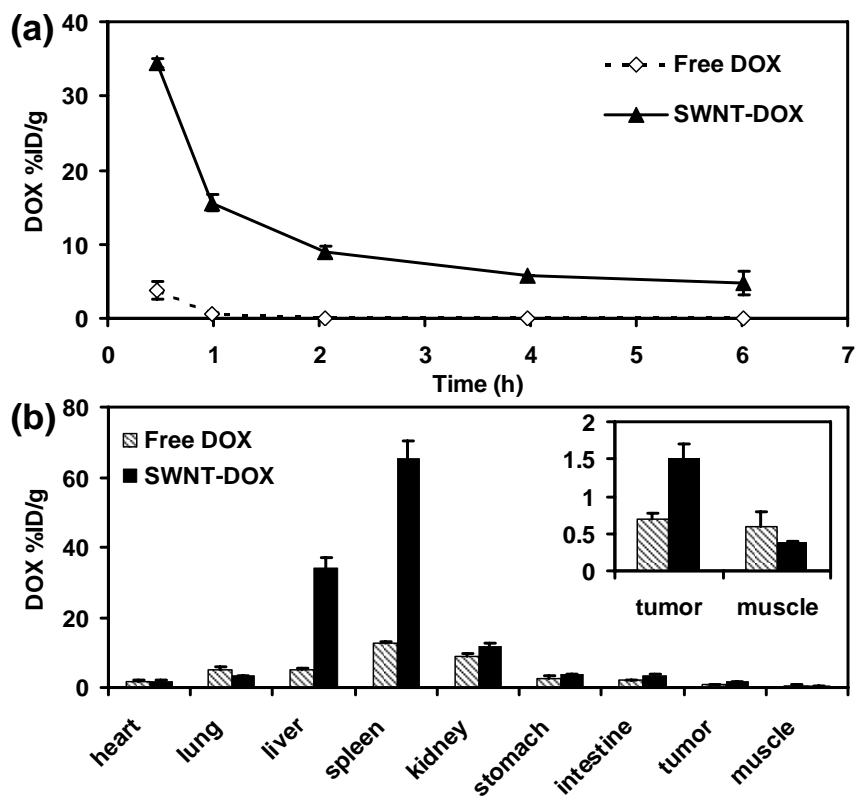

**Figure 2.** The pharmacokinetics and biodistribution of two DOX formulations were studied by fluorescence spectroscopy. **(a)** SWNT-DOX showed prolonged blood circulation compared with free DOX. DOX concentrations in blood from mice treated with free DOX and SWNT-DOX were measured by fluorescence spectroscopy at different time points post injection. **(b)** SWNT-DOX had higher tumor specific uptake and RES uptake than free DOX. Biodistribution of DOX in major organs of mice was measured 6 h after injection of free DOX and SWNT-DOX. Error bars were based on SEM of triplicate samples.

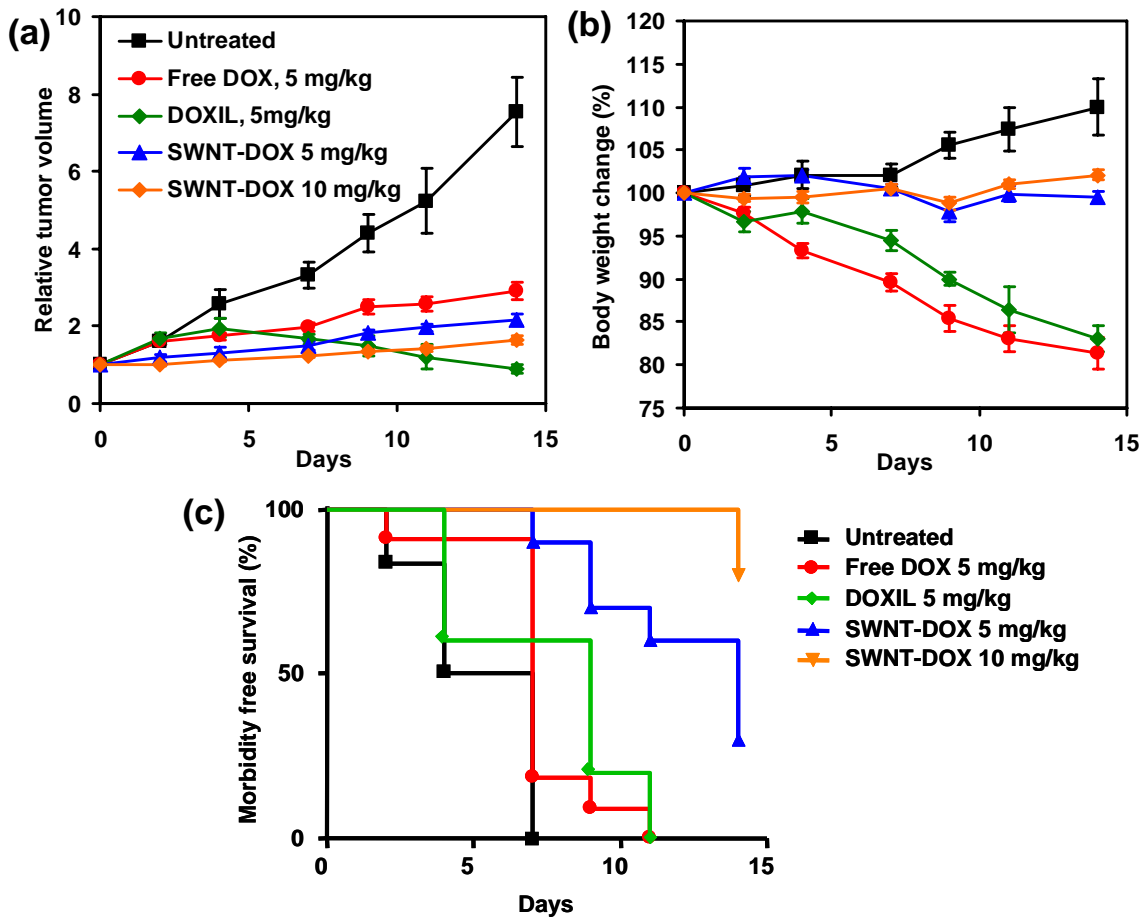

**Figure 3.** Supramolecular packing of DOX to SWNTs increased clinical efficacy *in vivo*. Raji tumor bearing SCID mice were treated with different DOX formulations once a week at day 0 and day 7. Tumor sizes and body weights were recorded three times a week. **(a)** SWNT-DOX formulation slowed tumor growth in a dose-dependent manner, showing better efficacy than free DOX but not DOXIL. Tumor sizes of untreated (n = 7), 5 mg/kg free DOX treated (n = 10, 2 mice died in the second week), 5 mg/kg Doxil treated (n = 5), 5 mg/kg SWNT-DOX treated (n = 10) and 10 mg/kg SWNT-DOX treated (n = 10) mice were measured during treatment. p values at 2 weeks: untreated *versus* DOX 5 mg/kg, $p = 0.001$; DOX 5 mg/kg *versus* SWNT-DOX 5 mg/kg, $p = 0.016$; DOX 5 mg/kg *versus* SWNT-DOX 10 mg/kg, $p < 0.001$; SWNT-DOX 5 mg/kg *versus* SWNT-DOX 10 mg/kg, $p = 0.018$. Mean tumor volume normalized to Day 0 +/- SEM is graphed. **(b)** SWNT-DOX resulted in far less weight loss than DOX and DOXIL. Mean body weight normalized to Day 0 +/- SEM is graphed. **(c)** SWNT-DOX formulation resulted in increased morbidity free survival. In the Kaplan-Maier analysis, we scored mice for morbidity when tumor volume increased by 2-fold, when body weight decreased by 10%, or when they died during treatment. p values: DOX 5 mg/kg *versus* SWNT-DOX 5mg/kg or 10mg/kg, $p < 0.001$; DOXIL 5mg/kg *versus* SWNT-DOX 5mg/kg, $p = 0.013$; DOXIL 5mg/kg *versus* SWNT-DOX 10mg/kg, $p < 0.001$)

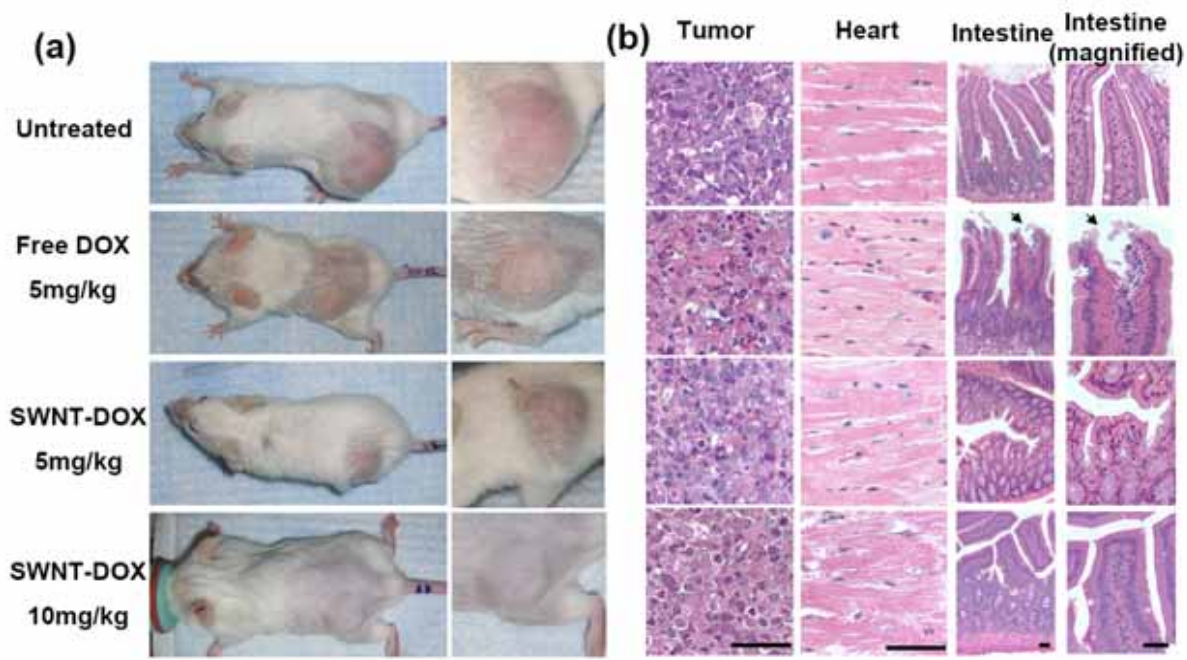

**Figure 4.** SWNT-DOX showed less *in vivo* toxicity. **(a)** Representative photos of mice from different groups were taken at the end of treatment. Mice became visibly thinner after treatment by free DOX. **(b)** Gastrointestinal toxicity was observed in free DOX treated mice but not in SWNT-DOX treated mice. Histological sections of intestinal epithelium showed damage of intestinal epithelium in the free DOX treated group. Arrows: area of loss of columnar epithelial cells in tips of villi. Scale bar: 100 microns.

# Supplementary information

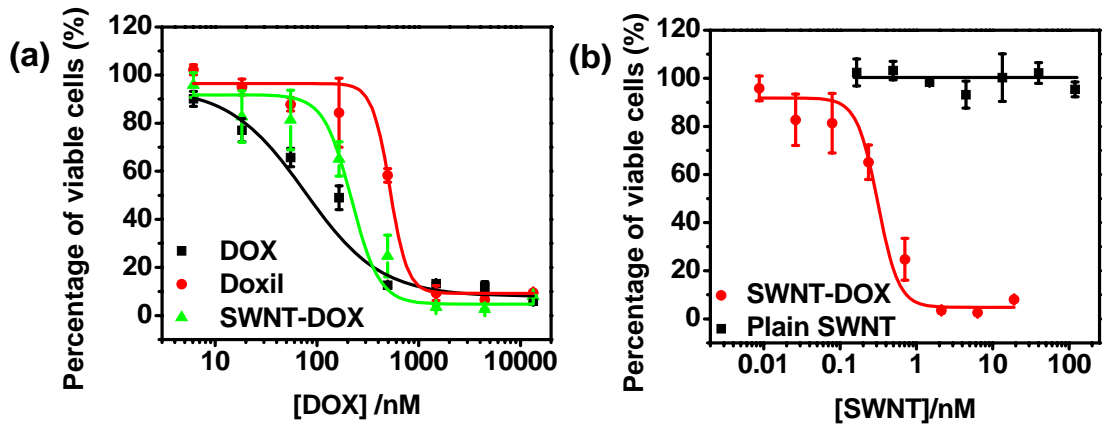

**Supplementary Figure S1.** SWNT-DOX was tumoricidal against lymphoma cells *in vitro*. **(a)** SWNT-DOX maintained cytotoxicity on human B-cell lymphoma Raji cells. Raji cells were incubated with free DOX, DOXIL and SWNT-DOX at series of concentrations for 3 days. Relative cell viabilities determined by MTS assay (*versus* untreated control) were plotted against DOX concentrations during cell incubation. IC$_{50}$ values were measured to be 76 ± 8.6 nM, 519 ± 11 nM and 218 ± 7.2 nM for free DOX, DOXIL and SWNT-DOX, respectively. **(b)** Plain SWNTs were non-toxic to Raji cells. Relative cell viabilities were plotted against SWNT concentrations (1 nM of SWNT corresponded to a weight concentration of 0.17 mg/L). While SWNT-DOX is highly toxic, plain SWNT exhibited no noticeable toxicity even at high concentrations after 3 days incubation with Raji cells. Error bars were based on standard deviations of triplicate samples.

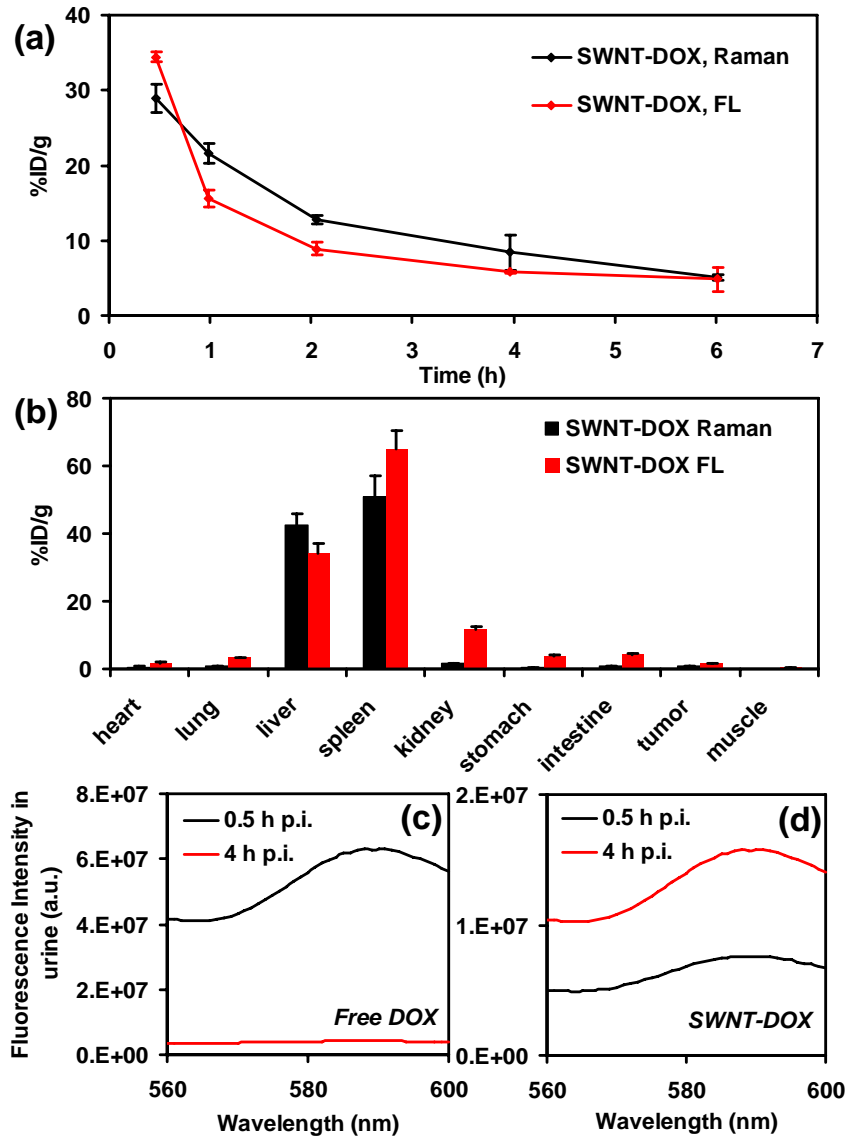

**Supplementary Figure S2.** *In vivo* behaviors of the SWNT-DOX complex. **(a)** Blood circulation of SWNT drug carrier and DOX was measured by Raman and fluorescence, respectively, in mice injected with SWNT-DOX. **(b)** Biodistribution of SWNT and DOX in mice injected with SWNT-DOX showed high RES uptake of both drug and drug carrier, indicating that DOX was carried into RES organs by SWNTs. **(c)** Free DOX was cleared from urine. The strong DOX fluorescence in the early urine sample (0.5 h) and low fluorescence at the later time point (4 h) indicated rapid renal clearance of free DOX. **(d)** SWNT-DOX had delayed urinary clearance of dissociated DOX. Increased DOX fluorescence was observed in the 4 h urine sample. Probing SWNTs ex vivo by Raman spectroscopy was based on our well established protocols published earlier.[1, 2] Error bars were based on SEM of triplicate samples.

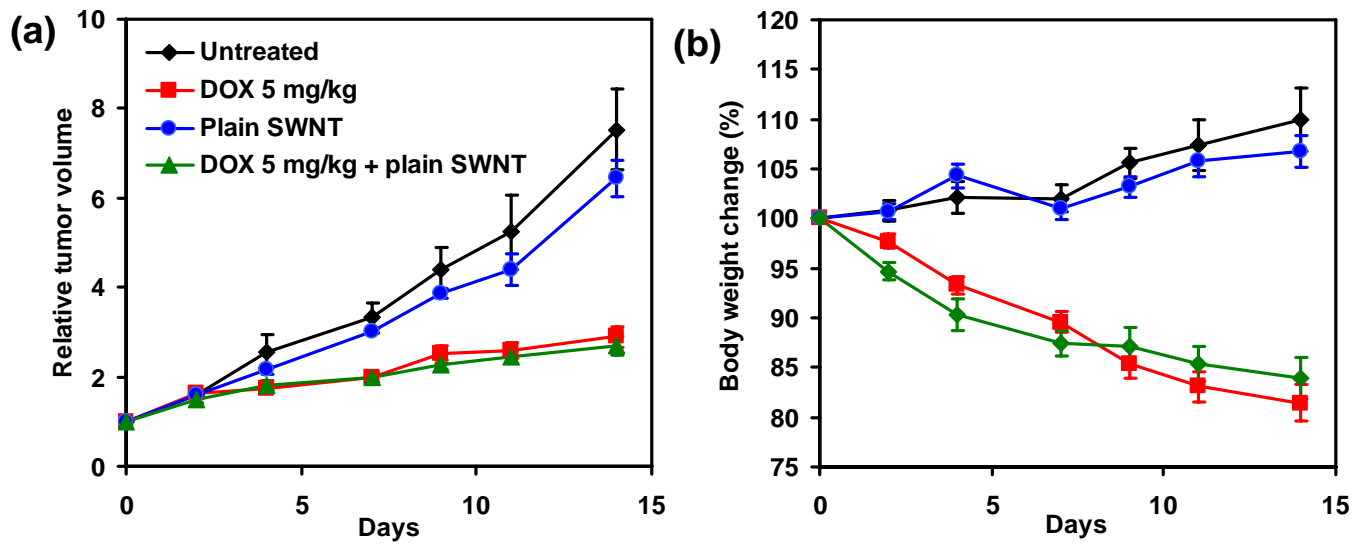

**Supplementary Figure S3.** Plain SWNTs exhibited no obvious effect on either tumor growth or mouse body weight change. Raji tumor bearing SCID mice were treated with plain SWNT, free DOX and free DOX plus plain SWNT once a week with tumor sizes and body weights recorded. The dose of plain SWNT was equal to that used in 5 mg/kg SWNT-DOX. **(a)** Plain SWNT had no *in vivo* tumoricidal effect. p values at 2 weeks: untreated *versus* plain SWNT, *p* = 0.34; DOX 5mg/kg *versus* DOX 5mg/kg + plain SWNT, *p* = 0.56. **(b)** Plain SWNT showed no noticeable effect on the mouse body weight change. p values at 2 weeks: untreated *versus* plain SWNT, *p* = 0.41; DOX 5mg/kg *versus* DOX 5mg/kg + plain SWNT, *p* = 0.32.

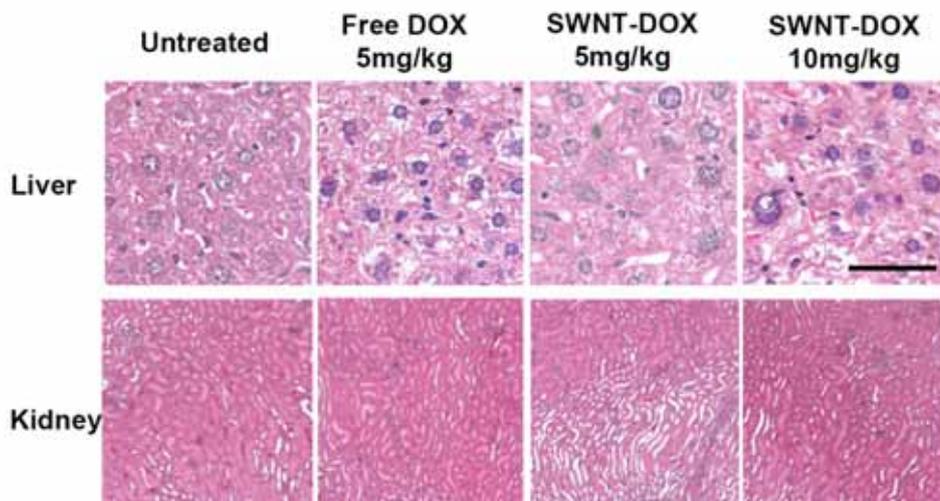

**Supplementary Figure S4.** Histological sections of liver and kidney from untreated, free DOX 5mg/kg, SWNT-DOX 5mg/kg and SWNT-DOX 10 mg/kg treated mice showed no obvious abnormality in those organs of various groups. Scale bar = 100 μm.

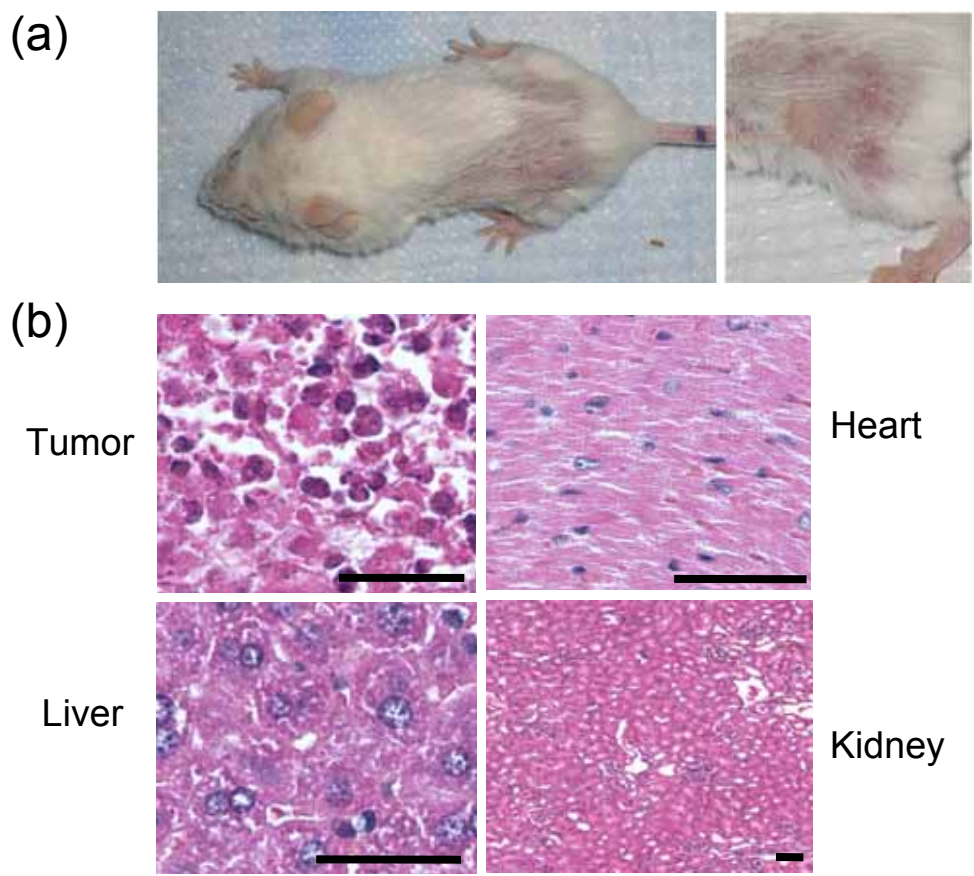

**Supplementary Figure S5.** Representative photos of the DOXIL treated group. **(a)** Representative photos of mice treated with DOXIL (5mg/kg) were taken at the end of treatment. **(b)** Histological sections of tumor, heart, liver and kidney from DOXIL (5mg/kg) treated mice. Toxicity caused by DOXIL treatment does not appear to affect these organs. Scale bar = 100 μm